# Semiconducting enriched carbon nanotube align arrays of tunable density and their electrical transport properties

Biddut K. Sarker[1,2], Shashank Shekhar[1,2] and Saiful I. Khondaker,[1,2,3*]

[1] Nanoscience Technology Center, [2] Department of Physics, [3] School of Electrical Engineering and Computer Science, University of Central Florida, 12424 Research Parkway, suite 400, Orlando, Florida 32826, USA

* To whom correspondence should be addressed. E-mail: saiful@ucf.edu

**Abstract**

We demonstrate assembly of solution processed semiconducting enriched (99%) single walled carbon nanotubes (s-SWNT) in an array with varying linear density via ac-dielectrophoresis and investigate detailed electronic transport properties of the fabricated devices. We show that (i) the quality of the alignment varies with frequency of the applied voltage and that (ii) by varying the frequency and concentration of the solution, we can control the linear density of the s-SWNTs in the array from 1/μm to 25 /μm. The maximum linear density of 25 s-SWNT /μm reported here is the highest for any aligned semiconducting array. The DEP assembled s-SWNT devices provide opportunity to investigate transport property of the arrays in the direct transport regime. Room temperature electron transport measurements of the fabricated devices show that with increasing nanotube density the device mobility increases while the current on-off ratio decreases dramatically. For the dense array, the device current density was 16 μA/μm, on-conductance was 390 μS, and sheet resistance was 30 kΩ/□. These values are the best reported so far for any semiconducting nanotube array.

**KEYWORDS:** Semiconducting carbon nanotube, aligned array, solution processed, dielectrophoresis, transistors,

Due to their exceptional electronic properties including very high mobility, near ballistic conductance and resistance against electromigration, single-walled carbon nanotubes (SWNTs) are considered to be a promising building block for future digital and analog electronics [1-2]. Devices fabricated from arrays of SWNTs can be advantageous as they average out device to device inhomogeneity of individual SWNTs and cover large areas. In addition, devices fabricated with nanotube arrays contain hundreds of parallel SWNTs contributing in charge transport, which can increase current outputs up to hundreds of microamperes. The degree of alignment and the density of SWNT in the array are expected to have a great influence on device performance [3-13]. For these reasons, there is a significant research effort to fabricate devices with massively parallel arrays of SWNTs for field effect transistors (FETs)[3-7], RF applications [9,10,14], plastic electronics [15,16], display technologies [17], electrodes [18,19]. and sensors[20].

The align array of SWNTs have been achieved using either direct growth via chemical vapor deposition (CVD) [5,9,10,21,22] or post growth from solution processed assembly techniques [3,4, 17-19,23-26]. However, such arrays contains a mixture of both semiconducting nanotubes (s-SWNT, 67%) and metallic nanotubes (m-SWNT, 33%), and the electric transport property of the arrays is dominated by the metallic pathways. In order to obtain better transistor properties, it is essential to selectively remove metallic nanotubes from the arrays via electrical breakdown [3-5, 27, 28]. Using this method, improved FET performance has been demonstrated; however, the device mobility was calculated without examining the number of nanotubes remaining in the device after electrical breakdown. In addition, in this method it is assumed that individual metallic nanotube can be removed selectively without affecting the neighboring nanotubes. Recent investigations reveal that removal of such a large number of nanotubes via electrical breakdown may have a detrimental effect on the remaining nanotubes in the array. For example, Wang et al have shown that although on-off ratio of FET was improved by electrical breakdown, the on current was decreased more (around 90%) than what was expected (33%) from breaking of metallic nanotubes [28]. Recently, we have shown that in a high density nanotube array, the electrical breakdown of one nanotube affects adjacent nanotubes due to the formation of dipole fields at the tip of broken nanotube. This leads to a highly correlated breakdown of neighboring nanotubes and semiconducting nanotubes are not immune to correlated breakdown [29]. Therefore, the removal of a large number of metallic nanotubes through electrical breakdown is a challenging and imperfect approach [14, 29, 30]. In this respect, it is important to fabricate devices consisting mostly of semiconducting nanotubes by minimizing or eliminating the metallic transport pathways where semiconducting enriched nanotubes hold the key.

Recently, it has been shown that solution based sorting technique such as density gradient ultracentifugation can provide highly enriched (99%) s-SWNTs in aqueous solution. A few studies have been reported on the FET fabrication using random network of such enriched s-SWNTs where transport occurs through the percolative pathways [31-33]. Compared to random networks where tube-tube junction limits charge transport, a perfectly aligned 2D array is expected to exhibit electronic properties that approached intrinsic properties of the individual nanotubes. Engel et al have used evaporation driven self assembly to obtain a 2D align s-SWNT array with a nanotube density of ~10 s-SWNT/µm which shown a sheet resistance of ~200 kΩ/□ and on-conductance of ~ 25 µS [25]. It is generally believed that the device property can be further improved if the density of the nanotube in the arrays can be increased further. In addition, it is also of great importance to investigate the electrical properties in the direct transport regime with varying density of s-SWNT for many practical applications.

In this paper, we demonstrate controlled assembly of semiconducting single walled carbon nanotubes array of tunable density from a high quality semiconducting enriched (99%) aqueous solution using AC dielectrophoresis (DEP). We show that the quality of the alignemt of the s-SWNT varies with frequency of the applied voltage. By varying the frequency and SWNT solution concentration, we can control the linear density of the nanotubes in the array from 1 s-SWNT/µm to 25 s-SWNT/µm. The maximum linear density of 25 s-SWNTs /µm reported here is the highest for any aligned s-SWNT array. The room temperature electron transport measurements of the fabricated FETs using the aligned array show that with increasing nanotube density, the device mobility increases while the current on-off ratio dramatically decreases. For the device with dense array, the current density is 16 µA/µm, on conductance is 390 µS, and sheet resistance is 30 kΩ/□. These values are the best reported so far for any semiconducting nanotube array. We also discuss possible reasons for low current on-off ratio at high linear density.

# RESULTS AND DISCUSSION

The nanotube assembly was done using dielectrophoresis (DEP) from a high purity (99%) s-SWNT aqueous solution purchased from NanoIntegris [34]. The diameter of the nanotubes varied from 0.5 to 3.9 nm with an average of 1.6 nm while the length of the nanotubes varied from 0.7 to 4.0 μm with an average value of 1.8 μm as measured by atomic force microscopy (AFM) (Figure 1a and b) (also see supporting information Figure S1). Figure 1c shows a schematic of the DEP assembly setup. The assembly was performed in a probe station between prefabricated palladium (Pd) source (S) and drain (D) electrodes of channel length $L = 2$ μm and channel width $W = 25$ μm. The original solution has a s-SWNT concentration of 10 μg

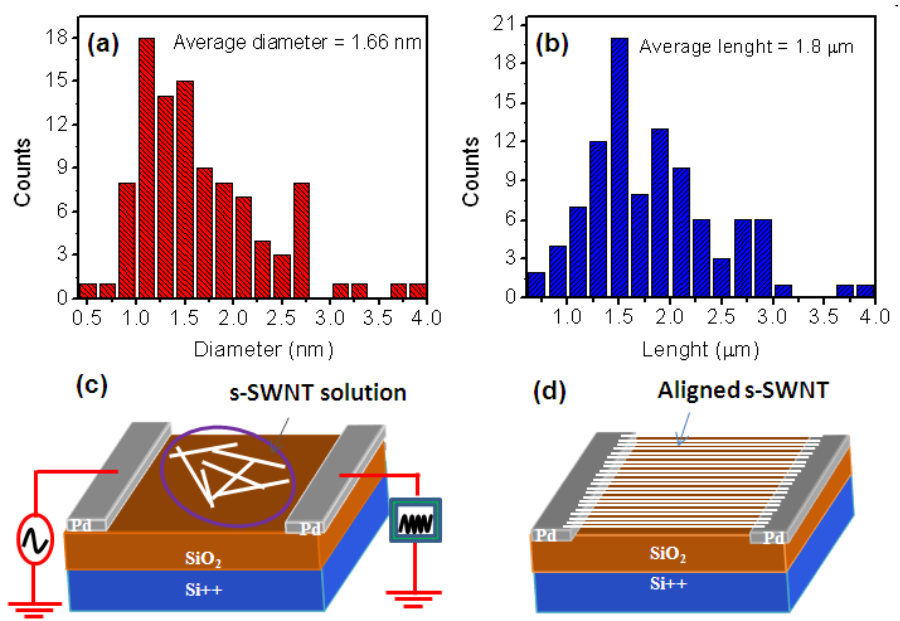

Figure 1. (a) Length distribution of the s-SWNT in the solution. The average length is 1.8 μm. (b) Diameter distribution of the s-SWNT. The average diameter is 1.6 nm. (c) Schematic of dielectrophoretic assembly of s-SWNT. A 3 μl of liquid droplet of s-SWNT solution was dispersed and an ac voltage of 5 V with varying frequency is applied between source and drain electrodes. (d) Schematic of nanotube alignment between palladium source and drain electrodes.

/ml and was further diluted to a desired concentration using de-ionized (DI) water (see details in material and methods section). For the assembly, a small (3 μL) drop of the s-SWNT solution was cast onto the chip containing electrode pairs. A fixed ac voltage ($V_{p-p}$) of 5 V was applied between source and drain electrodes for 40 s. The frequency was varied from 50 kHz to 5 MHz to study the quality of the assembled s-SWNT array. The ac field gives rise to a time averaged DEP force making the nanotubes move in a translational motion along the electric field gradient and align the nanotubes in the direction of the electric field lines (Figure 1d). We demonstrated in the past that DEP assembly technique is an effective approach to fabricate devices using 2D, 1D, and 0D nanomaterials (3, 4, 35-39). DEP can be advantageous over other solution processed techniques because it allows the materials to be directly integrated to prefabricated electrodes at the selected positions of the circuits and does not require post-etching or transfer printing.

After the DEP assembly, scanning electron microscope (SEM) was used to characterize the arrays. Figure 2 shows SEM images of s-SWNT assemblies at different frequencies for a fixed solution concentration of 100 ng/ml. It can be clearly seen from these images that the number of assembled nanotubes between the electrodes and the quality of the assembly strongly depends on the frequency of the applied voltage. At 50 kHz, a large number of short nanotubes are anchored to the electrode bases without bridging the electrode gap (Figure 2a). Although there are a few nanotubes that completely bridged the electrodes, however the alignment quality is poor. In addition some aggregation can also be seen. The quality of the assembly and their orientation is improved significantly when the frequency is increased to 300 kHz. This is shown in Figure 2b. In this assembly, a large number of individual nanotubes bridged the electrodes. This can be more clearly seen from the magnified images in Figure 2d. Majority of the nanotubes are reasonably aligned along with a few mis-aligned nanotubes. Although the nanotubes length varies from 0.7 μm to 4 μm in the solution, it is seen form the images that the most of the bridged nanotubes length are equal to the electrode gap. This suggests that the DEP process favors the alignment of nanotubes of length comparable to the electrode gap. One major drawback of the assembly at 300 kHz is that the distribution of the nanotube along the channel width is not uniform and some voids are seen. The more uniform distribution of nanotubes along with increased number of nanotubes in the assembly was achieved when the frequency was increased to 1 MHz (Figure 2c). A magnified view of this assembly is presented in Figure 2e. Although a majority of the nanotubes are aligned within ± $15^0$ of the longitudinal axis, some nanotubes were misaligned and went over the aligned nanotubes. Further increase in frequency to 2.5 MHz and 5 MHz (Figure 2g and 2h) resulted in a better alignment, however, the number of nanotubes bridging the electrodes decreased. In addition, some short nanotubes can be seen at the electrodes bases due to the electric field line alternation near nanotube-electrode junctions [3].

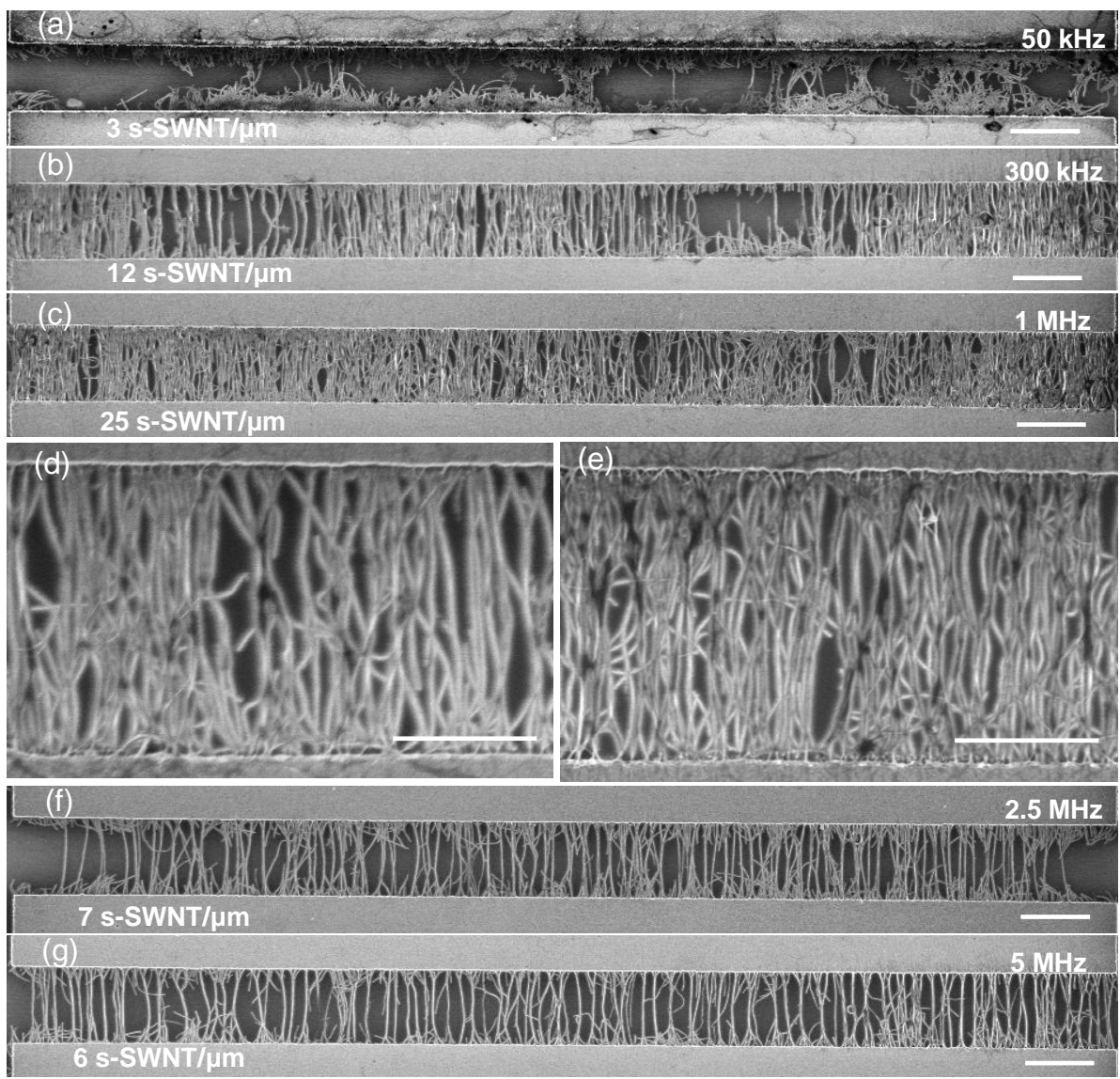

Figure 2. SEM image showing frequency dependence of s-SWNT assembly for a fixed solution concentration of 100 ng/ml at (a) 50 KHz (b) 300 KHz, (c) 1 MHz, (f) 2.5 MHz and (g) 5 MHz. Scale bars: 2 µm. (d, f) Magnified image of (c) and (d). Scale bars: 1 µm. The corresponding linear densities of the SWNTs in the array in Fig (a) –(d) are 3 s-SWNT/µm, 12 s-SWNT/µm, 25 s-SWNT/µm, 7 s-SWNT/µm and 6 s-SWNT/µm.

From these SEM images, we counted the total number of individual nanotubes that completely bridged the source and drain electrodes and divided the total by channel width to calculate the average linear density (*D*) of s-SWNT arrays. For the solution concentration of 100 ng/ml, the value of D at 50 kHz, 300 KHz, 1 MHz, 2.5 MHz and 5 MHz are 3 s-SWNT/μm, 12 s-SWNT/μm, 25 s-SWNT/μm, 7 s-SWNT/μm, and 6 s-SWNT/μm respectively. From these images, it is clear that with increasing frequency the quality of the alignment improves; however, the density of nanotubes initially increases with frequency and then decreases.

The variation of the nanotube density with frequency can be explained by DEP theory. Considering the nanotubes are cylindrical shape, the DEP force ($F_{DEP}$) exerted on a nanotube is given by [40], $F_{DEP} \propto \varepsilon_m \text{Re}[(\varepsilon_n^* - \varepsilon_m^*)/\varepsilon_m^*]\nabla E^2$; where $\varepsilon_{n,m}^* = \varepsilon_{n,m} - i\sigma_{n,m}/\omega$ with $\varepsilon_n$ and $\varepsilon_m$ are the dielectric constants of s-SWNT and deionized (DI) water respectively, σ is the conductivity, and *E* is the electric field. In low (*f*→0) and high (*f* →∞) frequency limit, the above mentioned equation takes the form $F_{DEP} \propto \varepsilon_m \text{Re}[(\sigma_n - \sigma_m)/\sigma_m]\nabla E^2$ and $F_{DEP} \propto \varepsilon_m \text{Re}[(\varepsilon_n - \varepsilon_m)/\varepsilon_m]\nabla E^2$ respectively. Here $\sigma_n$ is the conductivity of s-SWNT and $\sigma_m$ is the conductivity of DI water. At low frequency limit, $F_{DEP}$ is proportional to the difference of $\sigma_n$ and $\sigma_m$ while it is independent of both $\varepsilon_n$ and $\varepsilon_m$. It is reported that conductivity of s-SWNT is dominated by the surface conductivity rather than intrinsic conductivity [41]. Since our s-SWNT solution is ionic surfactant based (1% w/v), the conductivity of the solvent may be comparable to the intrinsic conductivity of s-SWNT. Therefore, s-SWNTs experience a weak $F_{DEP}$ at lower frequency. On the other hand, at high frequency limit, the $F_{DEP}$ depends on the difference of $\varepsilon_n$ and $\varepsilon_m$. The $\varepsilon_m$ is ~80 [42] and $\varepsilon_n$ is lies in between 2 to 5, depending on the band gap of s-SWNT [43]. So, $F_{DEP}$ becomes negative for s-SWNT at high frequency limit. From this discussion, it is clear that both at low frequency and high frequency limit, the DEP assembly process is not effective. As the frequency is increased, the DEP force will increase and then at some frequency it will start to decrease and become negative. This is consistent with our observation that the s-SWNT density initially increases with frequency and then decrease.

We have also investigated how the concentration of the s-SWNT solution affects the assembly. Figure 3 shows SEM images for the DEP assembled s-SWNT array for solution concentrations of 12.5, 25, 50 ng/ml and 100 ng/ml at a fixed frequency of 1 MHz. When the solution concentration is 12.5 ng/ml (Figure 3a), the average density is 1 s-SWNT/μm. Figure 3b and 3c shows SEM images of the array with solution concentrations of 25 ng/ml and 50 ng/ml respectively. It is seen from these images that the number of assembled nanotubes increased with solution concentration. The average densities are 6 s-SWNT/μm and ~ 18 s-SWNT/μm for concentration of 25 ng/ml and 50 ng/ml respectively. As the solution concentration is increased to 100 ng/ml (Figure 3c), we found that the linear density of nanotubes increased to ~25 s-SWNT/ μm. We have studied similar concentration dependent DEP assemblies at other frequencies of 50 KHz, 300 KHz, 2.5 MHz, and 5 MHz and the corresponding nanotubes densities were recorded from their respective SEM images (not shown here).

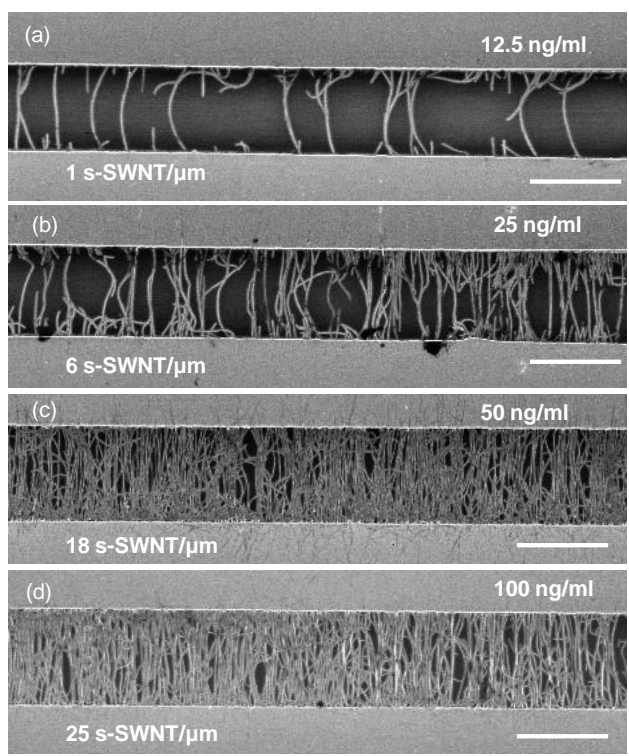

Figure 3. Solution concentration dependence of s-SWNT assembly at a fixed frequency of 1 MHz. SEM image show assembled nanotubes for solution concentration of (a) 12.5 ng/ml; (b) 25 ng/ml; (c) 50 ng/ml; (d) 100 ng/ml. Scale bar: 2 μm. The corresponding s-SWNT densities are 1 s-SWNT/μm, 6 s-SWNT/μm, 18 s-SWNT/μm and 25 s-SWNT/μm.

Figure 4 summarizes the DEP assembly of s-SWNT at different solution concentrations and different frequencies. Figure 4a shows the histogram of nanotube density with frequency at different solution concentrations. The maximum density of nanotubes is found at 1 MHz for all solution concentrations and it decreases both at lower and higher frequencies. Figure 4b shows how the density varies with solution concentrations. For a fixed frequency, the number of nanotubes increases with solution concentrations. From here, it can be seen that by tuning the frequency and solution concentration, we can vary the linear density of the s-SWNTs in the array from 1 s-SWNT/μm to 25 s-SWNT/μm. The maximum linear density of nanotube (25 s-SWNT/μm) reported here is the highest for the aligned semiconducting array.

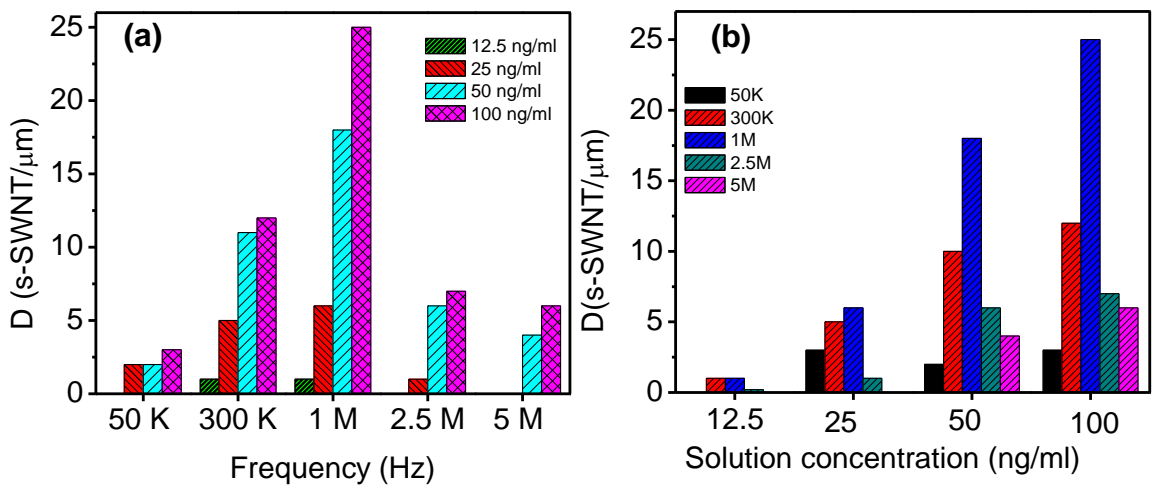

Figure 4. (a) The histogram showing variation of s-SWNT density with frequency at solution concentrations of 12.5 , 25, 50 and 100 ng/ml Maximum linear density is about 25 s-SWNT/μm in the frequency range 300KHz - 1MHz. Linear density of s-SWNT is decreased both at high and low frequencies. (b) Variation of density of s-SWNTs in the array with solution concentrations at frequency of 50 kHz, 300 kHz, 1 MHz, 2.5 MHz and 5 MHz.

To evaluate the electrical performance of the fabricated s-SWNT arrays with different nanotube densities, we have measured room temperature electronic transport properties in back-gated FET configuration. Figure 5a is a plot of drain current ($I_d$) as a function of source-drain voltage ($V_d$) of a representative device with 1 s-SWNT/μm at gate voltages $V_g$ = 0 V to -30 V with -5 V interval. The curve shows linear behavior with $V_d$, consistent with other short channel SWNT FET device[25] . In addition, the linear behavior of the $I_d$-$V_d$ curves at small $V_d$ bias ( -50 mV to 50 mV) indicates that the ohmic contact is formed between s-SWNTs and Pd electrodes. Figure 5b shows transfer characteristics ($I_d$ versus $V_g$ plot) of the same device at $V_d$ = -0.2 V to -1 V, with a -0.2 interval. The device shows p-type field effect behavior with a well-modulation with $V_g$. All the measured devices with different nanotube densities have shown similar output and transfer characteristics; however, the gate modulations as well as magnitude of their current values were different. Figure 5c exhibits the transfer characteristics of typical devices with nanotube density of 1, 5, 10, 20, and 25 s-SWNT/μm. All the devices were measured at a fixed $V_d$ = -1 V. As we see from this Figure that the $I_d$ of the device with 1 s-

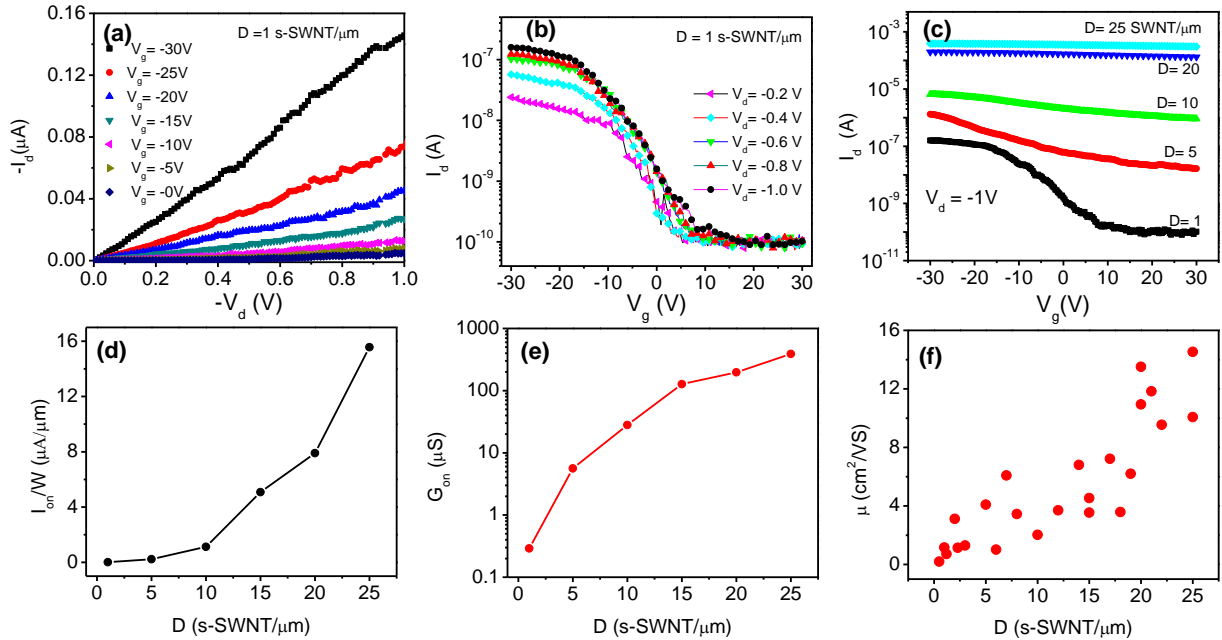

Figure 5: (a) Output characteristics ($I_d$-$V_d$) of a representative aligned array FET with linear density of 1 SWNT/μm at $V_g$ = 0 to -30V with – 5V interval. (b) Transfer characteristics ($I_d$-$V_g$) of the same device at $V_d$ = -0.2 V to -1 V with -0.2 V interval (c) $I_d$-$V_g$ curves at $V_d$= -1V of the aligned array FET devices with nanotube densities of 1, 5, 10, 20 and 25 s-SWNT/μm . (d) Variation of the normalized on current with nanotube densities. (e) Plot of on-conductance of the devices versus nanotube density in the array. (f) Mobility versus density of nanotubes of all the devices.

SWNT/μm is changed by several orders of magnitude with $V_g$ with a current on-off ratio of 1.5 ×10$^3$. However, as the nanotube density is increased in the devices, the current on-off ratio decreased. This will be discussed in more detail in Figure 6.

Another important observation in Figure 5c is that the on-current $I_{on}$ (at $V_g$=-30 V) increases with nanotube density. This can be more clearly seen in Figure 5d where we plot normalized on current ($I_{on}/W$) versus nanotube density, where $I_{on}$ was measured at $V_d$ = -1 V and $V_g$= -30 V. The maximum $I_{on}/W$= 15.6 μA/μm reported here is the highest reported for any semiconducting nanotube network and is a manifestation of remarkable properties of our dense s-SWNT array (25 s-SWNT/μm). The on state conductance ($G_{on} = I_{on}/V_d$) versus nanotube density plot (Figure 5e) shows $G_{on}$ also increases with nanotube density. In addition, it can be seen from this Figure that the device with 25 s-SWNTs/μm have a maximum $G_{on}$ of 390μS with a corresponding sheet resistance ($R_s$) of 30 kΩ/□. The maximum $G_{on}$ of our device is 16 times higher and the minimum $R_s$ values is 6 times lower than that of previous reported values of semiconducting nanotube array device with comparable device geometry ($L$= 4μm) [25]. The highest $G_{on}$, highest $I_{on}/W$ and lowest $R_s$ for our most dense array device are expected because our devices were fabricated with a higher nanotube density in the array (25 s-SWNT/μm) and with shorter channel length. In addition, the electronic transport occurs directly through individual s-SWNTs without any percolation through nanotube junctions.

Figure 5f shows the measured linear mobility ($\mu$) of all the devices as a function of nanotube density. The linear mobility is calculated using the standard formula $\mu=(L/WC_iV_d)(dI_d/dV_g)$, where, $dI_d/dV_g$ is the transconductance extracted from the slope of the transfer curve (see also supporting information Figure S2), $C_i = D/[C_Q^{-1} + (1/2\pi\varepsilon_0\varepsilon)\ln[\sinh(2\pi t_{ox}D)/\pi Dr]]$ is the specific capacitance per unit area of aligned array [4,5] with $C_Q$ is the quantum capacitance of nanotube (= 4×10$^{-10}$ F/m), $t_{ox}$ is the oxide thickness (= 250 nm); $\varepsilon$ is the dielectric constant of SiO$_2$ (3.9); $\varepsilon_0$ is the permittivity in the free space (8.85×10$^{-12}$ F/m) and $r$ is the averages radius of the nanotubes. It can be seen from Figure 5e that the mobility increases with nanotube density, similar to what has been reported for random network of s-SWNT [33].

In order to further investigate the effect of nanotube density on the device performance, the current on/off ratio $I_{on}/I_{off}$, the on-current $I_{on}$ and off-current $I_{off}$ of all the measured devices are plotted as a function of nanotube density in Figure 6. The plot of $I_{on}$ and $I_{off}$ with nanotube density (Figure 6a) shows that $I_{on}$ (red circle) increases with the nanotube

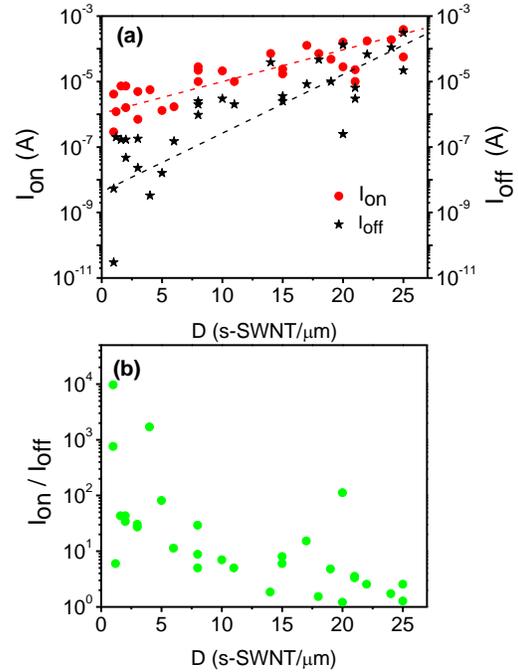

Figure 6: (a) On-current (red circle) and off-current (black star) variation as a function of number of SWNTs/μm in the array. (b) Plot of current on/off ratio versus number of SWNTs/μm.

density. On the other hand, $I_{off}$ (black star) is also increased with nanotube density. In addition, a large variation of $I_{off}$ is seen for the device with lower nanotube density. The plot of on/off ratio versus nanotube density (Figure 6b) demonstrates that devices with low density (~1 s-SWNTs/μm) exhibit a on/off ratio as high as ~$10^4$, although a large device to device variation from 7 to $10^4$ can be seen. However, the on-off ratio and their variations decrease significantly with increasing nanotube density. This may be due to several reasons. One possible reason could be the presence of a small number of m-SWNT in the array. The solution that was used for s-SWNT assembly contains 1% m-SWNT and it has been speculated in the past that m-SWNTs may be more favorable during DEP assembly process. In order to estimate the possible fraction of m-SWNT in our DEP assembled array, we have carried out DEP assembly of individual s-SWNTs in a taper shaped electrode design and measured their electrical transport properties. A representative SEM image of this assembly is shown in Figure 7a. Figure 7b is a $I_d$-$V_g$ curve of a

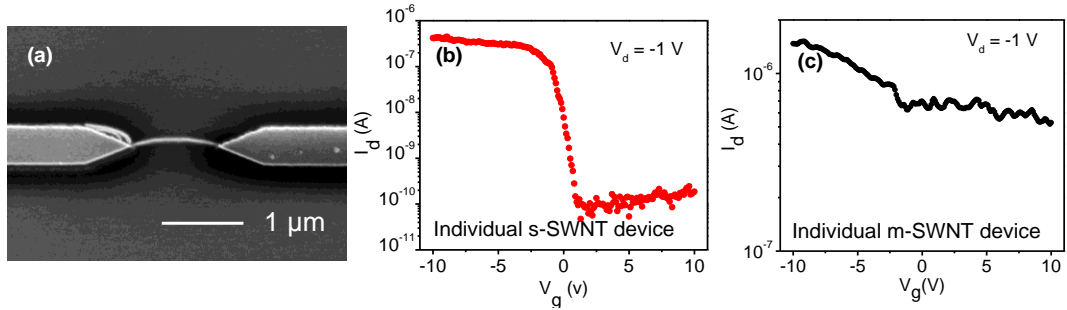

Figure 7: SEM image of an individual s-SWNT device. $I_d$-$V_g$ curve of a typical individual s-SWNT device (b) and m-SWNT device (c). The current on-off ratios of s-SWNT and m-SWNT device are $5\times10^3$ and 3 respectively. The bias voltage in the both curves was $V_d = -1$V.

typical individual s-SWNT device with on current ($I_{s, on}$) ~ $4\times10^{-7}$ A, off current ($I_{s, off}$) ~ $8\times10^{-11}$ A and current on-off ratio of $5\times10^3$. The $I_d$-$V_g$ curve of a typical m-SWNT presented in Figure 7c shows the on-off ratio of 3 with on current ($I_{m, on}$) ~$1.5\times10^{-6}$ A and off current ($I_{m, off}$) ~ $5\times10^{-7}$A. The details study of this individual s-SWNT assembly will be discussed elsewhere. The electrical measurement data of 87 such individual devices show that 85 nanotubes are semiconducting (on/off ratio ≥ 10) and 2 nanotubes are metallic (on/off ratio < 10). Using this individual assembly as a guide, we estimate that there may be up to 3% nanotubes in the array which could be metallic. In order to understand how this small percentage of m-SWNT can affect the property of the array, we estimated the on-off ratio of our aligned array using $I_{on}/I_{off} = (sI_{s,on} + mI_{m, on})/(sI_{s,off} + m I_{m, off})$, where $s$ and $m$ are the number of s-SWNT and m-SWNT respectively in the array. Table 1 summarizes the estimated on-off ratio of the FETs as function of probable number of m-SWNTs in the arrays. It can be seen from the table that for the array with 1 s-SWNT/um (total 25 s-SWNTs), if there is no m-SWNT present, the $I_{on}/I_{off}$ ratio is limited by the off current of s-SWNT, and $I_{on}/I_{off}$ could be 10000. However, if we consider that there may be two m-SWNTs, the $I_{on}/I_{off}$ ratio is decreased to 15. For the array with 1 s-SWNT/um (total 25 s-SWNTs) it is possible that some devices may contain no m-SWNT and some device may contain 1 or 2 m-SWNT, so a large variation in $I_{on}/I_{off}$ is not unreasonable. On the other hand, for the array of 25 s-SWNT/μm (total 625 s-SWNTs), it is almost certain that there may be a few m-SWNT. So the on-off ratio can vary from 106 to 30 for different

Table 1: Estimate of the current on-off ratio of the aligned array FETs as function of possible number of m-SWNTs in the array. The on current for s-SWNT and m-SWNT are taken $4\times10^{-7}$ A and $1.5\times10^{-6}$ A respectively. The off current for s-SWNT and m-SWNT are taken $8\times10^{-11}$ A and $5\times10^{-7}$ A respectively.

| 1 s-SWNT/μm = 25 s-SWNT | | 4 s-SWNT/μm = 100 s-SWNT | | 8 s-SWNT/μm = 200 s-SWNT | | 25 s-SWNT/μm = 625 s-SWNT | |
|---|---|---|---|---|---|---|---|
| *m-SWNT | On/off | *m-SWNT | On/off | *m-SWNT | On/off | *m-SWNT | On/off |
| 0 | 10000 | 1 | 102 | 2 | 102 | 6 | 106 |
| 1 | 26 | 2 | 52 | 4 | 52 | 12 | 54 |
| 2 | 15 | 3 | 35 | 6 | 35 | 18 | 30 |

Note: *m-SWNT denotes the most probable number of m-SWNT in the arrays

percentage of m-SWNTs. However, our experimental data at higher density array show that the on-off ratio is less than 10, which is not completely consistent with this model.

The possible reason for lower on-off ratio of the devices with higher nanotube densities could be due to the electrostatic screening effect of the gate voltage [44, 45]. A schematic illustration of this effect is shown in Figure 8. Our devices were fabricated on bare $SiO_2$ substrate in back gated configuration, and $t_{ox}$ and $L$ was 250 nm and 2 μm respectively. For the devices with lower density (~1 s-SWNT/μm), $t_{ox}$ is lower than the inter-nanotube spacing ($S$), and the electric field lines are not screened (Figure 8a). As a result, device with lower nanotube density is not affected by the screening [46, 47]. On the other hand, for the devices with higher nanotube density, $t_{ox}$ is larger than $S$ and the electric field lines of the nanotubes are screened by neighboring nanotubes (Figure 8c). This adversely affects the intrinsic capacitance of the nanotubes and hence the effect of the gate is reduced [44, 46, 47]. The gate voltage cannot deplete completely the charge carriers in the nanotubes which result the high off current and turns a lower on-off ratio. The critical density of our device is 4 s-SWNT/μm because at this density, $S$ is equal to $t_{ox}$ (Figure 8b). Therefore, on-off ratio of the device with density higher than 4 s-SWNT/μm should decrease significantly, which is consistent with our experiment finding (Figure 6b). The on-off ratio can be further improved by using a thinner gate oxide and work is in progress to that end.

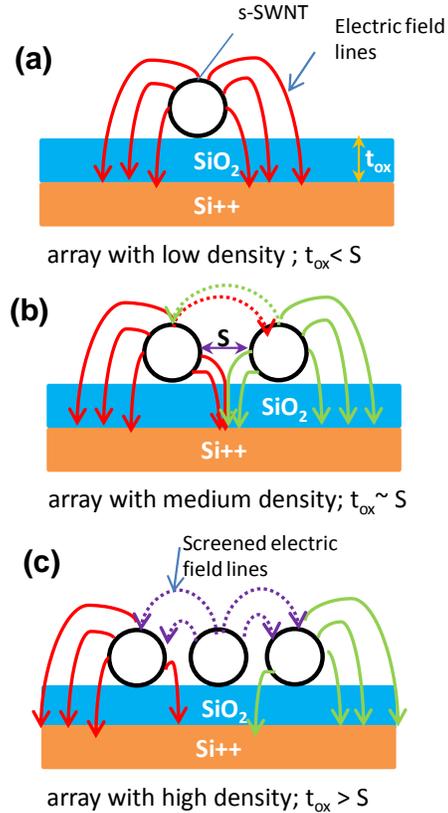

Figure 8: Schematic of the screening effect for different SWNT alignment densities: (a) Electric field lines of nanotubes when $t_{ox}$ is less than internanotube seperation (less SWNT density). No electric field is screened. (b) Electric field lines for medium density (here, 2 SWNTs are shown as the representative). Some of the electric field lines are screed due to neighbor nanotubes. (c) Electric field lines are screened at high nanotube density.

In summary, we have demonstrated for the first time that semiconducting nanotubes can be controllably aligned with a linear density from 1 s-SWNTs/μm to 25 s-SWNTs/μm by dielectrophoresis. The highest linear density (25 s-SWNTs /μm) achieved here is the maximum among the reported value for semiconducting assembly. The alignment densities were controlled independently by varying solution concentration and frequency of the applied electric field. At both low and high frequency limits, s-SWNT alignment density is low and it is higher in the frequency range 300 kHz -1 MHz. We found that the performance of the fabricated FETs is strongly influence by s-SWNT density in the device. The devices fabricated with highest density nanotube have shown the lowest sheet resistance, the highest on conductance and the highest on current density. We also observed that device mobility is increased with nanotube density and the highest device mobility 16 cm$^2$/Vs reported here is rather high considering that channel length (2 μm) of our devices. The current on/off ratio of the FET devices decreases with increasing nanotube density which can be attributed to the possible presence of a tiny fraction of m-SWNTs and the electrostatics screening effect. Our study has important implications for the large scale fabrication of FET devices with semiconducting carbon nanotubes.

**MATERIAL AND METHODS**

The 99% s-SWNT aqueous solution was purchased from NanoIntegris and it was free from catalytic particles and stable for at least six months.[31] The original solution has a s-SWNT concentration of ~ 10 μg/ml and it was diluted 100, 200, 400 and 800 times using de-ionized (DI) water to prepared concentration of 100, 50, 25 and 12.5 ng/ml respectively.

The directed assembly of s-SWNTs at predefined palladium electrodes was done in a probe station under ambient conditions via DEP. Pd electrodes were fabricated on heavily doped silicon substrates capped with a thermally grown 250 nm thick SiO$_2$ layer using standard EBL followed by deposition of Cr (3nm) and Pd (27 nm) and standard lift off. Prior to the assembly, the electrodes were placed in an oxygen plasma cleaner for 10 minutes to remove any unwanted organic residues on the surface. Pd was used because it is known to make the best electrical contact to SWNTs. The channel lengths (*L*) were 2 μm while the channel width (W) was 25 μm. For the assembly, a small (3 μL) droplet of the s-SWNT solution was cast onto the chip containing the 9 electrode pairs by a micropipet. A function generator (BK Prescision 4011A) was used to supply an AC voltage *of* $V_{p-p}$ =5V, with varying frequency of 50 KHz, 300 KHz, 1 MHz, 2.5 MHz and 5 MHz between the source and drain electrodes for *t* = 40 s. The signals were monitor by a dual trace oscilloscope (BK Prescision 2120B). After 40 s, the function generator was turned off and the remaining solution was blown out from the chip using a nitrogen gas flow.

The electrical transport measurements of the s-SWNT arrays were carried out by DL instruments 1211 current preamplifier and a Keithley 2400 source meter interfaced with LabView program. The SEM images of s-SWNT assembly were taken on Zeiss Ultra -55 SEM using Inlens detector with an accelerating voltage ~1 kV. Trapping mode AFM images were acquired using Dimension 3100 AFM (Veeco).

*Acknowledgment.* The help from Edwards Jimenez, Eliot Silbar and Narae Kang is greatly appreciated. This work is partially supported by the U.S. National Science Foundation under Grant ECCS-0748091 (CAREER).

*Supporting Information Available:* (1) AFM image and height profile of semiconducting nanotubes. (2) Variations of normalized transcendence with nanotube density. This material is available free of charge via internet at http://pubs.acs.org.

# Supporting Information

## Semiconducting enriched carbon nanotube align arrays of tunable density and their electrical transport properties

Biddut K. Sarker[1,2], Shashank Shekhar[1,2] and Saiful I. Khondaker,[1,2,3*]

[1] Nanoscience Technology Center, [2] Department of Physics, [3] School of Electrical Engineering and Computer Science, University of Central Florida, 12424 Research Parkway, suite 400, Orlando, Florida 32826, USA

* To whom correspondence should be addressed. E-mail: saiful@ucf.edu

1. **AFM image and height profile of semiconducting nanotubes**

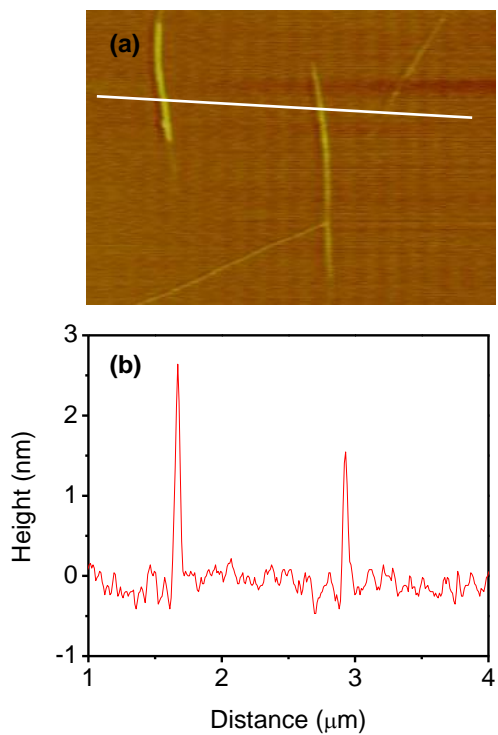

Figure S1: (a) Atomic force microscopy image of semiconducting carbon nanotubes (s-SWNT) dispersed on mica substrate. The s-SWNT solution was spin coated on the mica substrate at 1000 rpm for 60 s. (b) The height profile of the nanotubes. The diameters of nanotubes of these representatives individual s-SWNT are ~2.7 nm and ~1.8 nm. We also measured the length of the nanotubes by AFM. From such 100 measurements, we made the histogram of length and diameter distribution of nanotubes shown in Figure 1a and b.

## 2. Variations of normalized transcendence with nanotube density

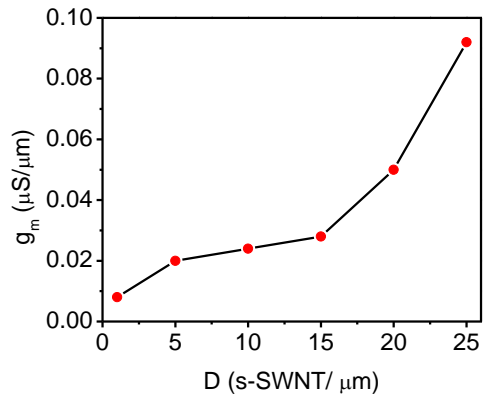

**Figure S2**: The variation of the normalized transconductance ($g_m/W$) with density of nanotubes in the device. In addition to on current density and on conductance, another important parameter figure of merit for transistor is transconductance ($g_m$). The $g_m$ is extracted from the slope of *I-Vg* curve and is normalized to device channel width (*W*). Figure shows that the ($g_m/W$) is increased with nanotube density. The maximum $g_m$ *and* maximum $g_m/W$ of the device with highest nanotube density (25 s-SWNT/μm) are found ~2.5 μS and ~0.1 μS/μm respectively.